# Superconducting properties of SmO$_{1-x}$F$_x$FeAs wires with T$_c$ = 52 K prepared by the powder-in-tube method

Zhaoshun Gao, Lei Wang, Yanpeng Qi, Dongliang Wang, Xianping Zhang, Yanwei Ma[*]

Key Laboratory of Applied Superconductivity, Institute of Electrical Engineering,
Chinese Academy of Sciences, P. O. Box 2703, Beijing 100190, China

Huan Yang, Haihu Wen

Institute of Physics, Chinese Academy of Sciences, Beijing 100190, China

**Abstract:**

We demonstrate that Ta sheathed SmO$_{1-x}$F$_x$FeAs wires were successfully fabricated by the powder-in-tube (PIT) method for the first time. Structural analysis by mean of x-ray diffraction shows that the main phase of SmO$_{1-x}$F$_x$FeAs was obtained by this synthesis method. The transition temperature of the SmO$_{0.65}$F$_{0.35}$FeAs wires was confirmed to be as high as 52 K. Based on magnetization measurements, it is found that a globe current can flow on macroscopic sample dimensions with $J_c$ of ~3.9×10$^3$ A/cm$^2$ at 5 K and self field, while a high $J_c$ about 2×10$^5$ A/cm$^2$ is observed within the grains, suggesting that a significant improvement in the globe $J_c$ is possible. It should be noted that the $J_c$ exhibits a very weak field dependence behavior. Furthermore, the upper critical fields ($H_{c2}$) determined according to the Werthamer-Helfand-Hohenberg formula are (T= 0 K) ≈ 120 T, indicating a very encouraging application of the new superconductors.

---

[*] Author to whom correspondence should be addressed; E-mail: ywma@mail.iee.ac.cn

The recent discovery of the iron based superconductor La(O$_{1-x}$F$_x$)FeAs with critical temperature 26 K has triggered great interest as a new family of high temperature superconductors [1]. Immediately, the $T_c$ has been enhanced up to above 50 K by the substitution of La using other smaller rare earth elements [2-7]. The present research indicates that the layered structure with the conducting Fe$_2$As$_2$ layers should be responsible for high temperature superconductivity，and the Re$_2$O$_2$ ( Re= rare earth ) layers behave as the charge reservoirs, all these look very similar to the case of cuprates. On the other hand, several groups have studied the $H_{c2}$, the irreversible magnetization and grain connectivity [8-15]. The resistive transition in fields suggests that $H_{c2}$ is remarkably high, which indicates encouraging potential applications. One crucial issue relevant for practical applications is the wire and tape fabrication. In particular, iron-based layered superconductors are mechanically hard and brittle and therefore not easy to drawing into the desired wire geometry. Recently, we have reported that Fe sheathed LaO$_{0.9}$F$_{0.1}$FeAs wires with Ti as a buffer layer were successfully fabricated by the powder-in-tube (PIT) method [16]. In this work, we have synthesized tantalum-clad SmO$_{01-x}$F$_x$FeAs wires and investigated the superconducting properties of SmO$_{1-x}$F$_x$FeAs wires.

The SmO$_{1-x}$F$_x$FeAs composite wires were prepared by the in situ powder-in-tube (PIT) method using Sm, As, SmF$_3$, Fe and Fe$_2$O$_3$ as starting materials. The raw materials were thoroughly grounded by hand with a mortar and pestle. The mixed powder was filled into a Ta tube of 8 mm outside diameter and 1 mm wall thickness. It is note that the grinding and packing processes were carried out in glove box in which high pure argon atmosphere is filled. After packing, the tube was rotary swaged and then drawn to wires of 2.25 mm in diameter. The wires were cut into 4~6 cm and sealed in a Fe tube. They are then annealed at 1160~1180 °C for 45 hours. The high purity argon gas was allowed to flow into the furnace during the heat-treatment process to reduce the oxidation of the samples.

The phase identification and crystal structure investigation were carried out using x-ray diffraction (XRD). Standard four probe resistance and magnetic measurements were carried out using a physical property measurement system (PPMS). Microstructure was studied using a scanning electron microscopy (SEM/EDX) after peeling away the Ta sheath.

Figure 1 shows the optical images for a typical transverse and longitudinal cross-section of $SmO_{1-x}F_xFeAs$ wires after heat treatment. A reaction layer with a thickness 10~30 μm was formed between the superconducting core and Ta tube due to the very high annealing temperature.

Figure 2 presents the XRD patterns of $SmO_{1-x}F_xFeAs$ wires after peeling off the sheath materials. As can be seen, the sample consists of a main superconducting phase, but some impurity phases are also detected. Such impurity phases might be reduced by optimizing the heating process and stoichiometry ratio of start materials. The lattice parameters are a = 3.9310(5) Å, c = 8.522(1) Å for the sample with x=0.35, and a = 3.9275(5) Å, c = 8.498(2) Å for the sample with x=0.3, which indicates a clear shrinkage of lattice parameters by increasing substitution of $F^-$ for $O^{2-}$.

Figure 3 shows the temperature dependence of resistivity at zero magnetic field for the samples $SmO_{1-x}F_xFeAs$. With decreasing temperature, the electrical resistivity of the sample with x=0.35 decreases monotonously and a rapid drop was observed starting at about 52 K, indicating the onset of superconductivity. The sample with x=0.30 shows the same behavior as that of the sample with x=0.35 except for a slightly low onset $T_c$. The residual resistivity ratio RRR = ρ(300K)/ ρ(55 K) were 2.8 and 2.3 for sample x=0.35 and x=0.3, respectively, indicating stronger impurity scattering than $LaO_{1-x}F_xFeAs$ [17].

Figure 4 shows the SEM images illustrating the typical microstructure of the fractured core layers for $SmO_{0.7}F_{0.3}FeAs$ wires. It can be seen that the $SmO_{0.7}F_{0.3}FeAs$ has a density structure with few voids in Fig. 4(a). EDX analysis also revealed that the sample possibly contains impurity phases such as SmAs and unreacted Fe or $Fe_2O_3$. The crystal structure of this kind of materials is based on a stack of alternating SmO and FeAs layers. A layer-by-layer growth pattern can be clearly seen as shown in Fig. 4(b), very similar to what has been observed in Bi-based cuprates.

The magnetization loops with a strong ferromagnetic background were observed in our samples, which are believed to be contributed by the unreacted Fe or $Fe_2O_3$ impurity phases [10]. We calculated the global $J_c$ for our samples on the basis of $J_c = 20\Delta M/a(1-a/3b)$, where $\Delta M$ is the height of the magnetization loop and $a$ and $b$ are the dimensions of the sample perpendicular to the magnetic field, $a < b$. Intragrain $J_c$ was also evaluated on the basis of $J_c = 30\Delta M/<R>$ through magnetic measurements after grinding the superconducting core into powder. $<R>$ is the average grain size, which is about 10 μm determined by SEM. The $J_c$ field dependence is shown in Fig.5. It can be seen that the $J_c$ based on the sample size is much lower than that what should exist in individual grains. The $J_c$ value at 5 K for the $SmO_{0.65}F_{0.35}FeAs$ powder is $2\times10^5$ A/cm$^2$ with a very weak dependence on field, which shows that the $SmO_{0.65}F_{0.35}FeAs$ has a fairly large pinning force in the grain. However, the global $J_c$ values of $3.9\times10^3$ A/cm$^2$ at 5 K obtained for the Sm samples is significantly lower than that seen in random bulks of $MgB_2$ which generally attained $10^6$ A/cm$^2$ at 4.2 K [18]. The main reason may be ascribed to the second impurity phases or the presence of weak links between the grains. As supported by XRD and SEM/EDX results, major impurity phases such as SmAs, SmOF, and unreacted Fe or $Fe_2O_3$ were observed in our present samples. These macroscale impurity phases can significantly reduce percolating current path and limit the globe $J_c$. Although the whole-sample current densities are significantly lower than that in random bulks of $MgB_2$, the grain connectivity seems better than random polycrystalline cuprates [12]. It is believed that significant improvement in the globe $J_c$ is expected upon either optimization of processing parameters or achieving high grain alignment in analogy to high $T_c$ Bi-based cuprates.

Figure 6 shows the resistive superconducting transitions for the $SmO_{0.65}F_{0.35}FeAs$ sample under magnetic fields. The large magnetoresistance at normal state may be caused by magnetic impurity. It is found that the onset transition temperature is not sensitive to magnetic field, but the zero resistance point shifts more quickly to lower temperatures due to the weak links or flux flow. We tried to estimate the upper critical field ($H_{c2}$) and irreversibility field ($H_{irr}$), using the 90% and 10% points on the resistive transition curves. The inset of Fig. 6 shows the temperature dependence of $H_{c2}$ and $H_{irr}$ with magnetic fields up to 9 T for the $SmO_{0.65}F_{0.35}FeAs$ wires. It is clear that the curve of $H_{c2}$ (T) is very steep with a slope of $-dH_{c2}/dT|_{T_c}$ = 2.74 T / K. For the type II superconductor in the dirty limit, $H_{c2}$ (0) is given by the Werthamer-Helfand-Hohenberg formula [19], $H_{c2}(0) = 0.693 \times (dH_{c2}/dT) \times T_c$. Taking $T_c$ = 51 K, we get $H_{c2}(0) \approx 96.8$ T. If adopting a criterion of 99 %$\rho_n$(T) instead of 90%$\rho_n$(T), the $H_{c2}(0)$ value of this sample obtained by this equation is higher than 120 T. We can also see that the irreversibility field in the inset of Fig. 5 is rather high comparing to that in $MgB_2$. These high values of $H_{c2}$ and $H_{irr}$ indicate that this new superconductor has an encouraging application in very high fields.

We have successfully prepared Ta clad $SmO_{1-x}F_xFeAs$ wires by the PIT process. The $J_c$ of composite wires has a very weak dependence on magnetic field with a much high $H_{c2}(0)$ above 120 T. Although some problems such as phase impurity remain unsolved, our preliminary results explicitly demonstrate that the feasibility of fabricating iron based superconducting wires.

The authors thank Xiaohang Li, Chenggang Zhuang and Liangzhen Lin for their help and useful discussion. We also thank Profs. Zizhao Gan, and Liye Xiao for the great encouragement. This work is partially supported by the Beijing Municipal Science and Technology Commission under Grant No. Z07000300700703, National '973' Program (Grant No. 2006CB601004) and National '863' Project (Grant No. 2006AA03Z203).

# Captions

Figure 1  Optical images for a typical transverse (a) and longitudinal (b) cross-section of the wires after heat treatment.

Figure 2  XRD patterns of $SmO_{1-x}F_xFeAs$ wires after peeling away the Ta sheath. The impurity phases of SmAs and SmOF are marked by * and #, respectively.

Figure 3  Temperature dependences of resistivity for $SmO_{1-x}F_xFeAs$ wires after peeling away the Ta sheath.

Figure 4  (a) Low magnification and (b) high magnification SEM micrographs for the $SmO_{0.7}F_{0.3}FeAs$ samples.

Figure 5  Magnetic field dependence of $J_c$ at 5 K for the bar and powder of $SmO_{1-x}F_xFeAs$ wires.

Figure 6  Temperature dependence of resistivity under magnetic fields up to 9 T for the $SmO_{0.65}F_{0.35}FeAs$ wire. The inset shows the temperature dependence of $H_{c2}$ and $H_{irr}$ determined from 90% and 10% points on the resistive transition curves.

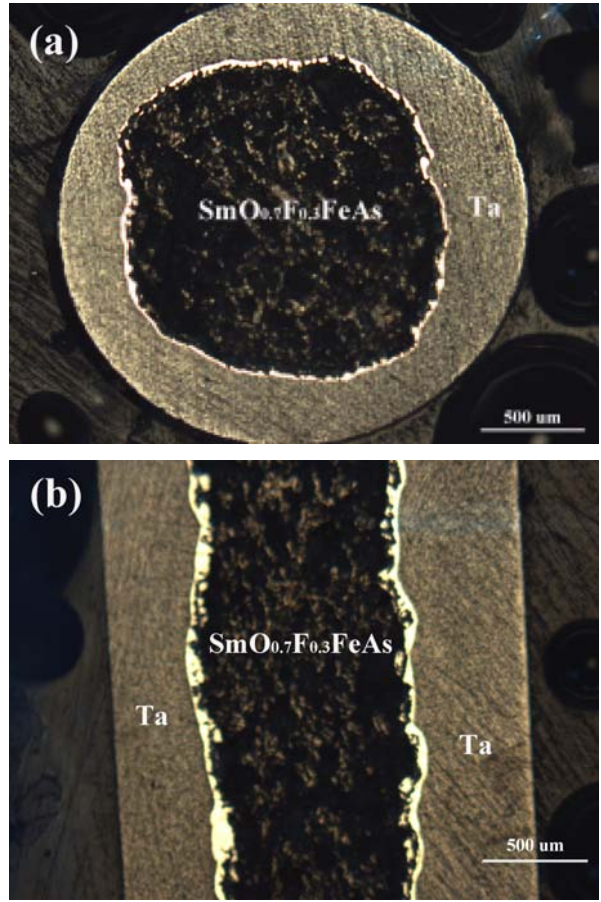

Fig.1 Gao et al.

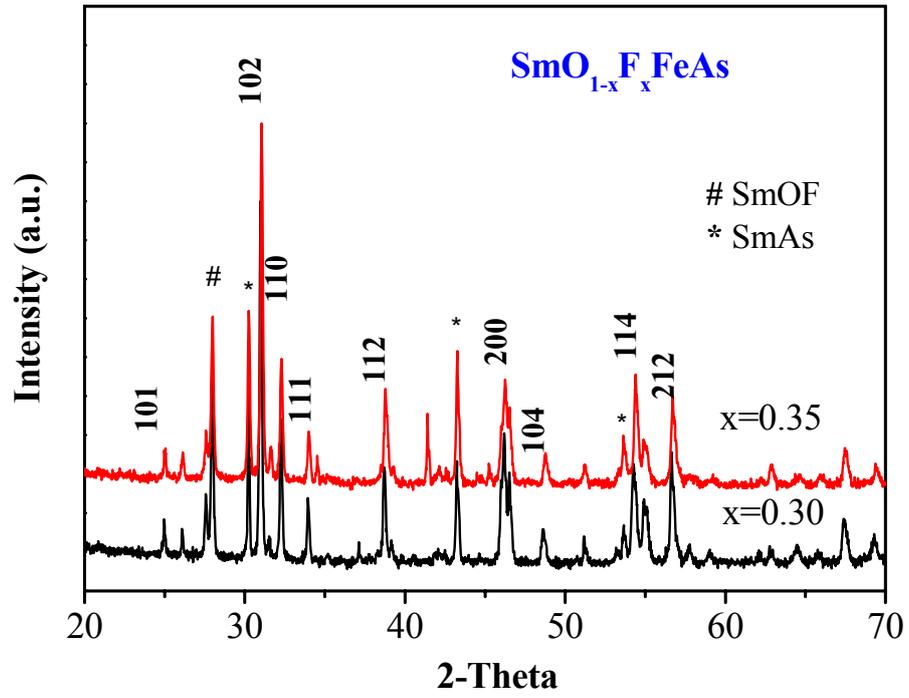

Fig.2 Gao et al.

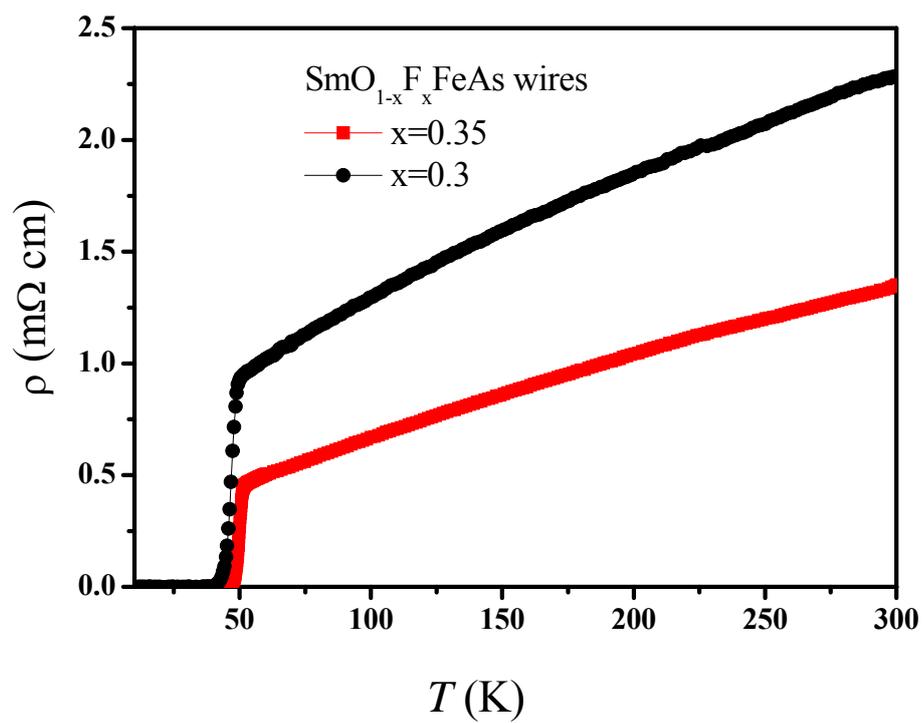

Fig.3 Gao et al.

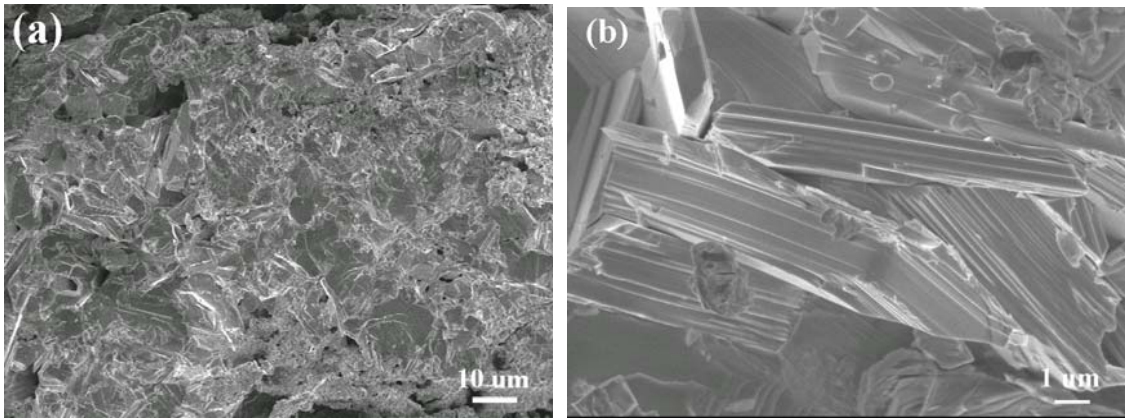

Fig.4 Gao et al.

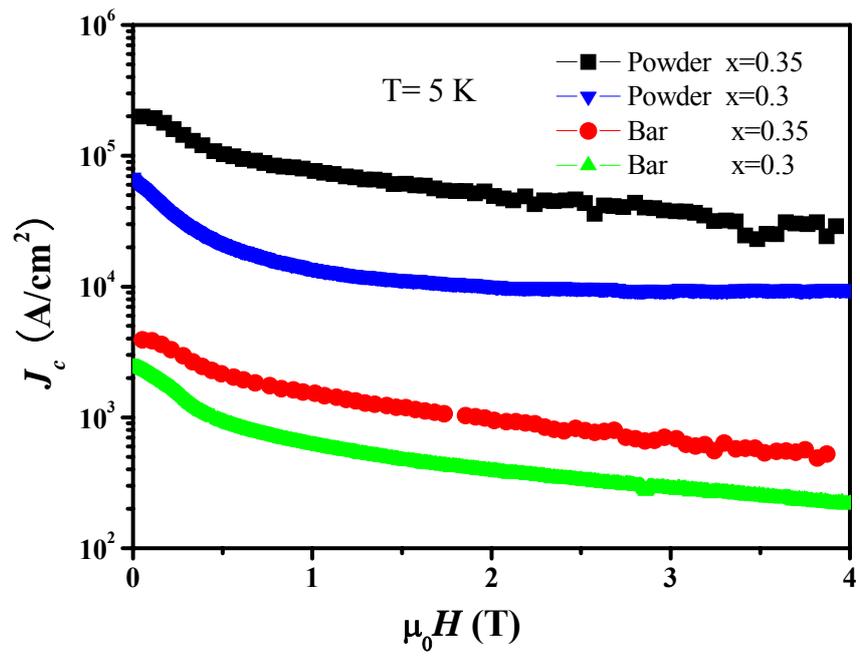

Fig.5 Gao et al.

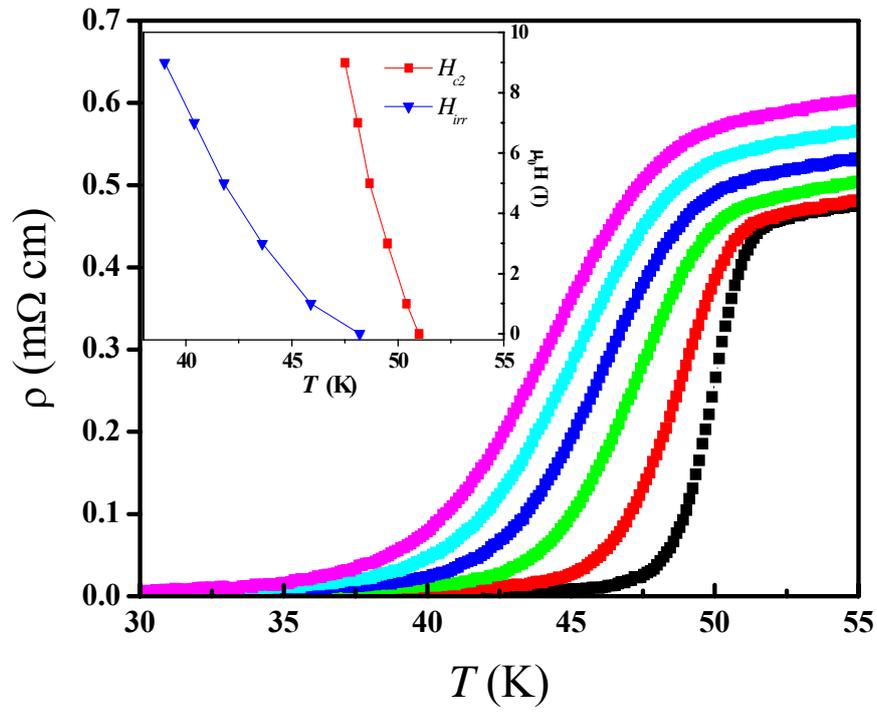

Fig.6 Gao et al.